\documentclass[preprintnumbers,amsmath,amssymb]{revtex4}

\usepackage{epsfig}
\usepackage{rotating}
\usepackage{graphicx}% Include figure files
\usepackage{dcolumn}% Align table columns on decimal point
\usepackage{bm}% bold math

\begin{document}

\title{
Semi-supervised learning by search of optimal target vector}

\author{Leonardo Angelini, Daniele Marinazzo, Mario Pellicoro and Sebastiano Stramaglia}

\affiliation{ TIRES-Center of Innovative Technologies for Signal Detection
and Processing, \\ Universit\`a di Bari, Italy \\
Dipartimento Interateneo di Fisica, Bari, Italy \\
Istituto Nazionale di Fisica Nucleare, Sezione di Bari, Italy
}

% The \author macro works with any number of authors. There are two commands
% used to separate the names and addresses of multiple authors: \And and \AND.
%
% Using \And between authors leaves it to \LaTeX{} to determine where to break
% the lines. Using \AND forces a linebreak at that point. So, if \LaTeX{}
% puts 3 of 4 authors names on the first line, and the last on the second
% line, try using \AND instead of \And before the third author name.

\date{\today}

\begin{abstract}
We introduce a semi-supervised learning estimator which tends to the
first kernel principal component as the number of labeled points
vanishes. Our approach is based on the notion of optimal target
vector, which is defined as follows. Given an input data-set of
${\bf x}$ values, the optimal target vector $\mathbf{y}$ is such
that treating it as the target and using kernel ridge regression to
model the dependency of $y$ on ${\bf x}$, the training error
achieves its minimum value. For an unlabeled data set, the first
kernel principal component is the optimal vector. In the case one is
given a partially labeled data set, still one may look for the
optimal target vector minimizing the training error. We use this new
estimator in two directions. As a substitute of kernel principal
component analysis, in the case one has some labeled data, to
produce dimensionality reduction. Second, to develop a
semi-supervised regression and classification algorithm for
transductive inference. We show application of the proposed method
in both directions.
\end{abstract}

\maketitle

\section{Introduction}
\label{intro} The problem of effectively combining {\it unlabeled} data with {\it
labeled} data, semi-supervised learning, is of central importance in machine learning;
see, for example, \cite{zu,zhu,ch} and references therein. Semi-supervised learning
methods usually assume that adjacent points and/or points in the same structure (group,
cluster) should have similar labels; one may assume that data are situated on a low
dimensional manifold which can be approximated by a weighted discrete graph whose
vertices are identified with the empirical (labeled and unlabeled) data points. This can
be seen as a form of regularization \cite{smo}. A common feature of these methods, see
also \cite{arg}, is that, as the number of labeled points vanishes, the solution tends to
the constant vector. An interesting survey on semi-supervised learning literature may be
found on the web \cite{zz}. Improving regression with unlabeled data is the problem
considered in \cite{zhou}, where co-training is achieved using k-NN regressors. A
statistical physics approach, based on the Potts model, is described in \cite{getz}. An
issue closely related to semi-supervised learning is active-learning: some attempts to
combine active learning and semi-supervised learning has been made \cite{zzz}.

The purpose of this work is to introduce a semi-supervised learning estimator which, as
the number of labeled points vanishes, tends to the first kernel principal component
\cite{kpca}; when a suitable number of labeled points is available, it may be used for
transductive inference \cite{vapnik}. Our approach is based on the following fact. Given
an unlabeled data set, its first kernel principal component is such that, treating it as
target vector, supervised kernel ridge regression provides the minimum training error.
Now, suppose that you are given a partially labeled data set: still one may look for the
target vector minimizing the training error. This optimal target vector may be seen as
the generalization of the first kernel principal component to the semi-supervised case.

The paper is organized as follows. In the next Section we describe our approach, while in
Section 3 the experiments we performed are described. Some conclusions are drawn in
Section 4.

\section{Methods} \label{methods}
\subsection{Kernel ridge regression}
We briefly recall the properties of kernel ridge regression (KRR), while referring the
reader to \cite{st} for further technical details. Let us consider a set of $\ell$
independent, identically distributed data $S=\{ ({\bf x}_i, y_i) \}_{i=1}^\ell$, where
${\bf x}_i$ is the $n$-dimensional vector of input variables and $y_i$ is the scalar
output variable. Data are drawn from an unknown probability distribution; we assume that
both ${\bf x}$ and $y$ have been centered, i.e. they have been linearly transformed to
have zero mean. The regularized linear predictor is $y={\bf w}\cdot{\bf x}$, where ${\bf
w}$ minimizes the following functional:
\begin{equation}\label{lagrangian-function-without-b}
L(\mathbf{w}) =  \sum_{i=1}^\ell \left ( y_i - {\bf w}\cdot {\bf x}_i \right )^2 +
\lambda || {\mathbf w} ||^2.
\end{equation}
Here $|| \mathbf{w} || = \sqrt{\bf{w}\cdot\bf{w}} $ and $\lambda >0$ is the
regularization parameter. For $\lambda  =0$, predictor
(\ref{lagrangian-function-without-b}) is invariant when new variables, statistically
independent of input and target variables, are added to the set of input variables (IIV
property, \cite{as}). One may show that this invariance property holds, for
(\ref{lagrangian-function-without-b}), also at finite $\lambda
> 0$.

KRR is the {\it kernel} version of the previous predictor. Calling $\bf{y}$ = $(y_1, y_2,
..., y_\ell)^\top$ the vector formed by the $\ell$ values of the output variable and
 $K(\cdot,\cdot)$  being a positive definite symmetric function, the predictor has the
following form:
\begin{equation}\label{notlinear}
y = f({\bf x}) = \sum_{i=1}^\ell c_i K({\bf x}_i,{\bf x}),
\end{equation}
where coefficients $\{c_i\}$ are given by
\begin{equation}\label{w2}
\bf{c} =  \left (\bf{K} + \lambda \bf{I} \right)^{-1} \bf{y},
\end{equation}
$\bf{K}$ being the $\ell \times \ell$ matrix with elements $K({\bf x}_i,{\bf x}_j)$.
Equation (\ref{notlinear}) may be seen to correspond to a linear predictor in the feature
space $ \Phi({\bf x}) = ( \sqrt{\alpha_1}\psi_1 ({\bf x}), \sqrt{\alpha_2}\psi_2 ({\bf
x}), ...,\sqrt{\alpha_N} \psi_N ({\bf x}), ... ), $ where $\alpha_i$ and $\psi_i$  are
the eigenvalues and eigenfunctions of the integral operator with kernel $K$. One may
 show \cite{prep} that, for KRR predictors with nonlinear kernels, the IIV property does not
generically hold, even for those kernels, discussed in \cite{as}, for which the property
holds at $\lambda=0$. Regularization breaks the IIV invariance in those cases.

Due to (\ref{notlinear}) and (\ref{w2}), the predicted output vector $\bf{\bar{y}}$, in
correspondence of the {\it true} target vector $\bf{y}$, is given by
$\bf{\bar{y}}$$=\mathbf{G}\bf{y}$, where the symmetric matrix $\mathbf{G}$ is given by
\begin{equation}\label{G}
\mathbf{G}=\mathbf{K}\left(\mathbf{K}+\lambda \mathbf{I}\right)^{-1}.\end{equation} Note
that matrix $\mathbf{G}$ depends only on the distribution of $\{\mathbf{x}\}$ values:
$\mathbf{G}$ embodies information about the structures present in $\{\mathbf{x}\}$ data
set. Indeed, for $i\ne j$, the matrix element $G_{ij}$ quantifies how much the target
value of the $j-th$ point influences the estimate of the target of point $i$. Let us now
consider the leave-one-out scheme; let data point $i$ be removed from the data set and
the model be trained using the remaining $\ell -1$ points. We denote $\tilde{y}_i$ the
target value thus predicted, in correspondence of $\bf{x_i}$. It is well known \cite{st}
that the leave-one-out-error $\tilde{y}_i -y_i$ and the training error obtained using the
whole data set $\bar{y}_i -y_i$ satisfy:
\begin{equation}
\label{loo} \tilde{y}_i -y_i={\bar{y}_i -y_i \over 1-G_{ii}}.
\end{equation}
This formula shows that the closer $G_{ii}$ to one, the farther the leave-one-out
predicted value from those obtained using also point $i$ in the training stage. Consider
a point $i$ in a dense region of the feature space:  one may expect that removing this
point from the data-set would not change much the estimate since it can be well predicted
on the basis of values of  neighboring points. Therefore points in low density regions of
the feature space are characterized by diagonal values $G_{ii}$ close to one, while
$G_{ii}$ is close to zero for points $\mathbf{x_i}$ in dense regions: the diagonal
elements of $\mathbf{G}$ thus convey information about the structure of points in the
feature space. It is worth stressing that, given a kernel function, the corresponding
features $\psi_\gamma ({\bf x})$ are not centered in general. One can show \cite{kpca}
that centering the features ($\psi_\gamma \to \psi_\gamma -\langle \psi_\gamma\rangle$,
for all $\gamma$) amounts to perform the following transformation on the kernel matrix:
$$\mathbf{K }\to \mathbf{\tilde{K}}=\mathbf{K}-\mathbf{I}_\ell \mathbf{K}-\mathbf{K}\mathbf{I}_\ell +\mathbf{I}_\ell \mathbf{K} \mathbf{I}_\ell,$$
where $\left( I_\ell\right)_{ij}=1/\ell$, and to work with the centered kernel
$\mathbf{\tilde{K}}$. In the following we will  assume that the kernel matrix
$\mathbf{K}$ has been centered.
\subsection{Optimal target vector}
The training error of the KRR model  is proportional to
$(\bf{y}-\mathbf{G}\bf{y})^\top(\bf{y}-\mathbf{G}\bf{y})=\bf{y}^\top \mathbf{H}\bf{y},$
where $\mathbf{H}= \mathbf{I}-2\mathbf{G}+\mathbf{G}\mathbf{G}$ is a symmetric and
positive matrix. In the unsupervised case the data set is made of $\mathbf{x}$ points,
$\{ {\bf x}_i\}_{i=1}^\ell$, the target function $\mathbf{y}$ is missing. However we may
pose the following  question: what is the vector $\mathbf{y}\in \mathbf{R}^\ell$ such
that treating it as the target vector leads to the best fit, i.e. the minimum training
error $\bf{y}^\top \mathbf{H}\bf{y}$? We expect that this {\it optimal} target vector
would bring information about the structures present in the data. To avoid the trivial
solution $\mathbf{y}=\mathbf{0}$, we constrain the target vector to have unit norm,
$\mathbf{y}^\top\mathbf{y}=1$; it follows that the optimal vector is the normalized
eigenvector of $\mathbf{H}$ with the smallest eigenvalue. On the other hand, matrix
$\mathbf{H}$ is a function of matrix $\mathbf{K}$: hence it has the same eigenvectors of
$\mathbf{K}$ while the corresponding eigenvalues $\mu_H$ and $\mu_K$ are related by the
following monotonically decreasing correspondence:
$$\mu_H=\left(1-{\mu_K\over \mu_K+\lambda}\right)^2.$$ Therefore,
independently of $\lambda$,  the smallest eigenvalue of $\mathbf{H}$ corresponds to the
largest eigenvalue of $\mathbf{K}$, and the optimal vector coincides with the first
kernel principal component. To conclude this subsection, we have shown that the method in
[10] may be motivated also as the search for the optimal target vector.

The notion of optimal target vector has been introduced in \cite{ang}, where a kernel
method for dichotomic clustering has been proposed, consisting in finding the ground
state of a class of Ising models.

\subsection{Semi-supervised learning}
Now we consider the case that we are given a set $S=\{ {\bf x}_i \}_{i=1}^\ell$ of data
points with unknown targets $\{t_i\}_{i=1}^\ell$, and a set $S'=\{ ({\bf x}_j, u_j)
\}_{j=\ell+1}^{N}$, where $N=\ell +m$, of input-output data. Without loss of generality
we assume that the labeled points belong to two classes, and take $u_j\in
\{-1/\sqrt{N},+1/\sqrt{N}\}$ for all $j$'s. The $N$ dimensional full vector of targets
$\mathbf{y}$ is obtained appending $\{t\}$ (unknown) and $\{u\}$ (known) values:
$$\mathbf{y}=(\mathbf{t}^\top \mathbf{u}^\top)^\top.$$
Keeping  the kernel and $\lambda$ fixed, we look for the unit norm target vector
$\mathbf{y}$ minimizing the training error $\mathbf{y}^\top \mathbf{H} \mathbf{y}$.  The
$N\times N$ matrix $\mathbf{H}$ has the block structure
\[ \mathbf{H} = \left( \begin{array}{cc}
              \mathbf{H_0} & \mathbf{H_1} \\
              \mathbf{H_1^\top}& \mathbf{H_2}
    \end{array}\right), \]
where $\mathbf{H_0}$ is an $\ell \times \ell$ matrix. Neglecting a constant term, the
optimal vector is determined by the vector $\mathbf{t}$ minimizing
\begin{equation}
\mathcal{E}(\bf{t})=\mathbf{t}^\top \mathbf{H_0} \mathbf{t} +2\mathbf{t}^\top
\mathbf{H_1} \mathbf{u} \label{eeee}
\end{equation}
 under the constraint $|| \mathbf{t} ||^2=1-|| \mathbf{u} ||^2$.
The first term of $\mathcal{E}$ favors projections of the $\ell$ points with great
variance, whereas the second term measures their consistency with  labeled points. Let us
denote $\{\Psi_{\alpha'}\}$ and $\{\mu_{\alpha'}\}$ the eigenvectors and eigenvalues of
$\mathbf{H_0}$, sorted into increasing $\mu_{\alpha'}$. We express
$\mathbf{t}=\sum_{\alpha' =1}^\ell \xi_{\alpha'} \Psi_{\alpha'}$. The coefficients
$\xi_{\alpha'}$ for the minimum are given by
$$\xi_{\alpha'} ={f_{\alpha'} \over \mu -\mu_{\alpha'}},$$
where $f_{\alpha'}= \Psi_{\alpha'}^\top\mathbf{H_1} \mathbf{u}$, and $\mu$ is a Lagrange
multiplier which must to be tuned to satisfy:
\begin{equation} \label{csi} g(\mu)= \sum_{\alpha' =1}^\ell
\left({f_{\alpha'} \over \mu -\mu_{\alpha'}}\right)^2=1-|| \mathbf{u} ||^2.
\end{equation}

Equation (\ref{csi}) has always at least one solution with $\mu < \mu_1$, see figure 1,
and usually this is the one minimizing $\mathcal{E}$. However all the solutions of
(\ref{csi}) must be compared according to their {\it energies} $\mathcal{E}$; those
corresponding to the lowest $\mathcal{E}$, $\bf{y^\star}$, is then  selected. Clearly as
$m\to 0$ one recovers the first eigenvector of $\mathbf{H_0}$, i.e. the first kernel
principal component: $\bf{y^\star}$ thus constitutes a generalization of the latter to
the semi-supervised case. To construct the other generalized kernel principal components,
we make the following transformation on matrix $\mathbf{H}$:
$$\mathbf{\tilde{H}}=\mathbf{H}-\mathbf{P^\star}\mathbf{H}
-\mathbf{H}\mathbf{P^\star}+\mathbf{P^\star}\mathbf{H}\mathbf{P^\star},$$ where
$\mathbf{P^\star}=\bf{y^\star}\bf{y^\star}^\top$ is the projector on the linear subspace
spanned by $\bf{y^\star}$. The symmetric matrix $\mathbf{\tilde{H}}$ has   the lowest
eigenvalue equal to zero and corresponding to eigenvector $\bf{y^\star}$. The system of
eigenvectors of $\mathbf{\tilde{H}}$ constitutes a generalization of kernel principal
components to the semi-supervised case. \section{Experiments}
\subsection{Generalizing kernel principal components}
Now we present some simulations of the proposed method, focusing on the dimensionality
reduction issue and comparing  with  fully unsupervised kernel principal component
analysis. We consider three well known data sets: IRIS (100 points in a four-dimensional
space, second and third classes, versicolor and virginica); colon cancer data set of
\cite{alon}, consisting in 40 tumor and 22 normal colon tissues samples, each sample
being described by the $100$ most discriminant genes; the leukemia data set of
\cite{golub}, consisting of samples of tissues of bone marrow samples, $47$ affected by
acute myeloid leukemia (AML) and $25$ by acute lymphoblastic leukemia (ALL), each sample
being described by the $500$ most discriminant genes. The following question is
addressed: is $\bf{y^\star}$ more correlated to the true labels than the fully
unsupervised first kernel principal component? Here we restrict our analysis to the
linear kernel.

We start with IRIS and proceed as follows. We randomly select $m=4$ points and, treating
them as labeled, we find the system of eigenvectors of $\mathbf{\tilde{H}}$. Then  we
evaluate the linear correlation $R$ between the eigenvectors and the true labels of the
whole data-set. The distributions of $R$ for the four eigenvectors are depicted in figure
2. We observe that in most cases the vector $\bf{y^\star}$ is more correlated with the
true classes than the fully unsupervised principal component: the one-dimensional
projection of data onto $\bf{y^\star}$ is more informative than the first principal
component. However there are situations where use of labeled points leads to poor
results; a typical example is depicted in figure 3. In figure 4 a situation is depicted
where knowledge of labeled points leads to a relevant improvement.

In general, we denote $f$ the fraction of instances such that $\bf{y^\star}$ is more
correlated to the true labels than the first principal component. In figure 5 we depict
$f$ as a function of $\bar{m}=m/N$ for the three data sets here considered. At
$\bar{m}=0.16$  $f$ is already nearly one. The semi-supervised method here proposed
outperforms principal components almost always for large $\bar{m}$.

\subsection{Transductive inference}
In this subsection we demonstrate the effectiveness of the proposed approach for
estimating the values of a function at a set of test points, given a set of input-output
data points, without estimating (as an intermediate step) the regression function.

The boston data set is a well-known problem where one is required to estimate house
prices according to various statistics based on $13$ locational, economic and structural
features from data collected by U.S. Census Service in the Boston Massachusetts area. For
$\ell =5,10,15,20,25$, we partition the data-set of $N=506$ observations randomly 100
times into a training set of $N-\ell$ observations and a testing set of $\ell$
observations. We use a Gaussian kernel with $\sigma =1$ and set $\lambda =1$; results are
stable against variations of these parameters. In Table 1 we report the mean squared
error (MSE) on the test set averaged over the 100 runs, for each value of $\ell$, we
obtain using the optimal target vector $\bf{y^\star}$. In Table 1 we also report the MSE
obtained using the classical KRR in the two step procedure: (i) estimation of the
regression function using the training data-set (ii) calculation of the regression
function at points of interest (test data-set). The improvement achieved using the
optimal target approach, over classical KRR, is clear.
\begin{table}
\caption{\label{tab:table1}The mean square error on the Boston data set obtained using
the optimal target (OT) approach and the classical kernel ridge regression (KRR) method.
The size of the test set is $\ell$. }
\begin{ruledtabular}
\begin{tabular}{lcr}
$\ell$&OT&KRR\\
\hline
    5&   2.3790   & 3.6312\\

   10&   2.7938  &  4.0111 \\

   15& 2.9460 &   4.1057 \\

   20&   3.1024 &   4.1802 \\

   25&   3.1569 &   4.1653 \\
\end{tabular}
\end{ruledtabular}
\end{table}

We also consider five well known data sets of pattern recognition from UCI database: we
evaluate the optimal target vector, points are then attributed to classes according to
the sign of $\bf{y^\star}$. We compare with the transductive linear discrimination (TLD)
approach developed in \cite{trans}; the performance of a classifier is measured by its
average error over 100 partitions of the data-sets into training and testing sets. We use
the linear kernel with $\lambda =1$, however the results are stable to variations of
$\lambda$. Obviously, our approach and TLD are applied to the same partitions of
data-sets, so that the comparison is meaningful. The results are shown in Table 2: our
approach outperforms TLD.
\begin{table}
\caption{\label{tab:table2}The percentage test error of transductive linear
discrimination and optimal target approach, on five datasets from UCI database.}
\begin{ruledtabular}
\begin{tabular}{lcr}
&TLD&OT\\
\hline
    Diabetes&   23.3   & 11.98\\

   Titanic&   22.4  &  6.52 \\

   Breast Cancer& 25.7 &   16.7 \\

   Heart&   15.7 &   3.3 \\

   Thyroid&   4.0 &   4.0 \\
\end{tabular}
\end{ruledtabular}
\end{table}

It is worth stressing that our results are obtained without a fine-tuning of parameters.
In particular,note that our definition of optimal target vector fixes the relative
importance of the two terms in equation (\ref{eeee}).
\section{Conclusions}
\label{conc} We have presented a new  approach to semi-supervised learning based on the
notion of optimal target vector, the target vector such that KRR provides the minimum
training error over all the possible target vectors. The proposed algorithm is
characterized by the fact that the first kernel principal component is recovered as the
cardinality of labeled points vanishes; hence it may be seen as a semi-supervised
generalization of Kernel Principal Components Analysis. The effectiveness of the proposed
approach for transductive inference has also been demonstrated.

\vskip 0.4 cm\par\noindent{\bf Acknoledgements.} The authors thank Olivier Chapelle for a
valuable correspondence on the subject of this paper. Discussions on semi-supervised
learning with Eytan Domany and Noam Shental are warmly acknowledged.

\begin{figure}[ht!]
\begin{center}
\epsfig{file=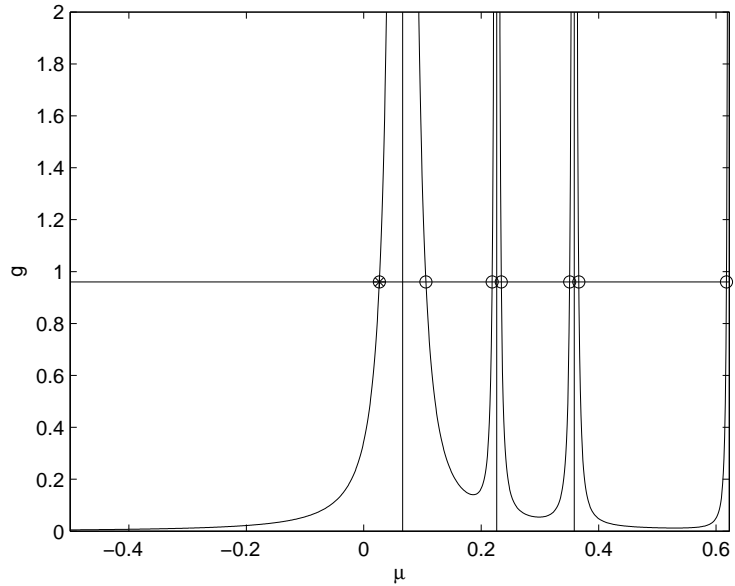,height=8.5cm}
\end{center}
\caption{{\small  The solutions of equation (\ref{csi}) are depicted, for a typical
instance  of four labeled points in the IRIS data set. The star corresponds to the
solution with $\mu < \mu_1$, which has the smallest energy $\mathcal{E}$.\label{fig1}}}
\end{figure}
\begin{figure}[ht!]
\begin{center}
\epsfig{file=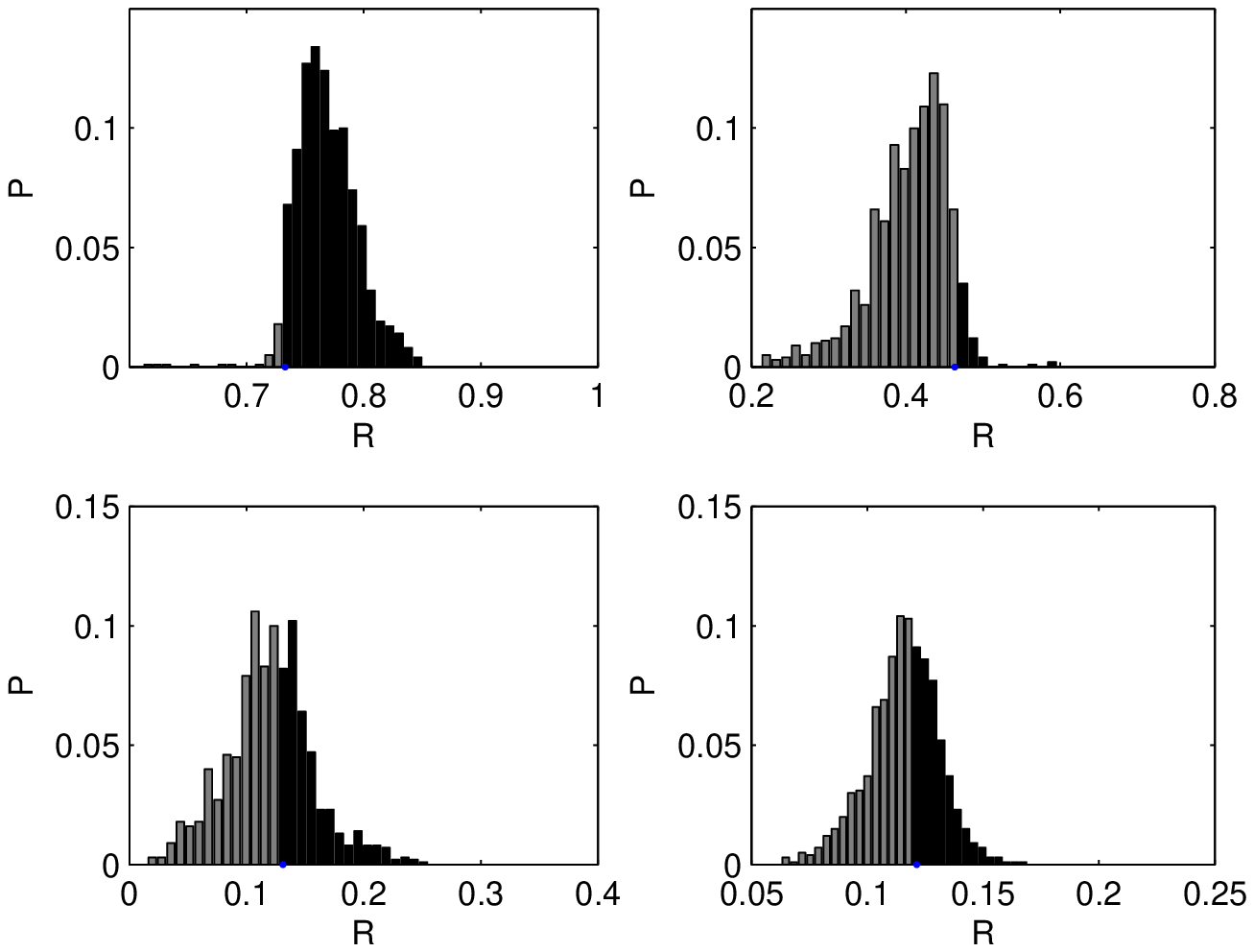,height=8.5cm}
\end{center}
\caption{{\small  Concerning IRIS data set and $m=4$, we depict the distribution (over
10000 random selections of labeled points) of the linear correlation $R$  between
eigenvectors of $\mathbf{\tilde{H}}$ and the true labels. From the left to the right and
the top to the bottom, we refer to the first, the second, the third and the fourth
eigenvector. Grey (black) histogram bars denote values of $R$ lower (greater) than those
of the corresponding fully unsupervised principal component. \label{fig2}}}
\end{figure}

\begin{figure}[ht!]
\begin{center}
\epsfig{file=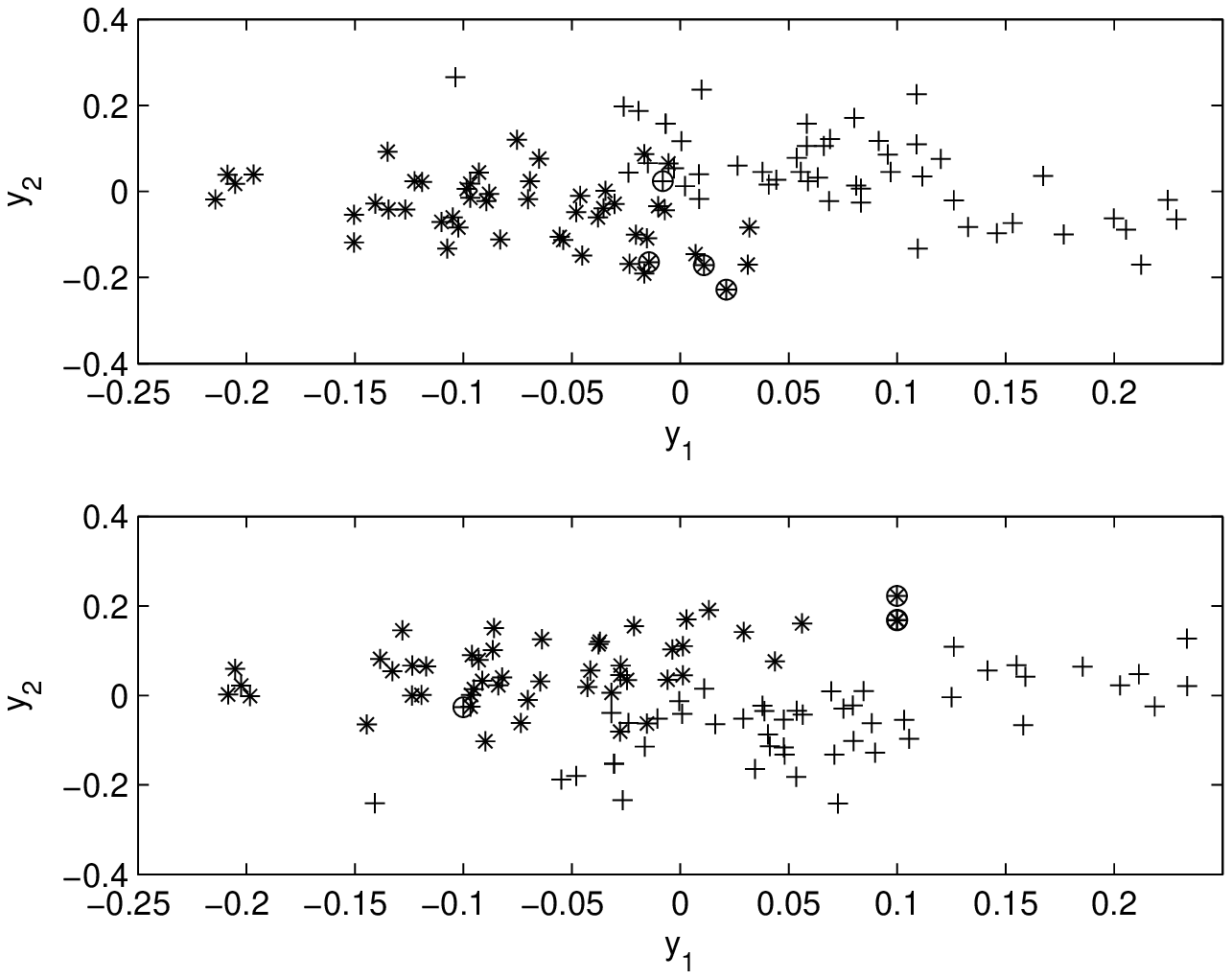,height=8.5cm}
\end{center}
\caption{{\small  (Top) The IRIS data set is depicted in the plane of the first two
principal components, $\star$ versicolor, $+$ virginica. The linear correlation of the
first principal component with the true labels is $R=0.732$. Four selected points are
surrounded by a circle. (Bottom) The data set is represented in the plane of the first
two eigenvectors of $\mathbf{\tilde{H}}$. The linear correlation between $\bf{y^\star}$
and  the true labels is $R=0.615$. (Note that two circles are almost overlapping and thus
difficult to distinguish). \label{fig3}}}
\end{figure}

\begin{figure}[ht!]
\begin{center}
\epsfig{file=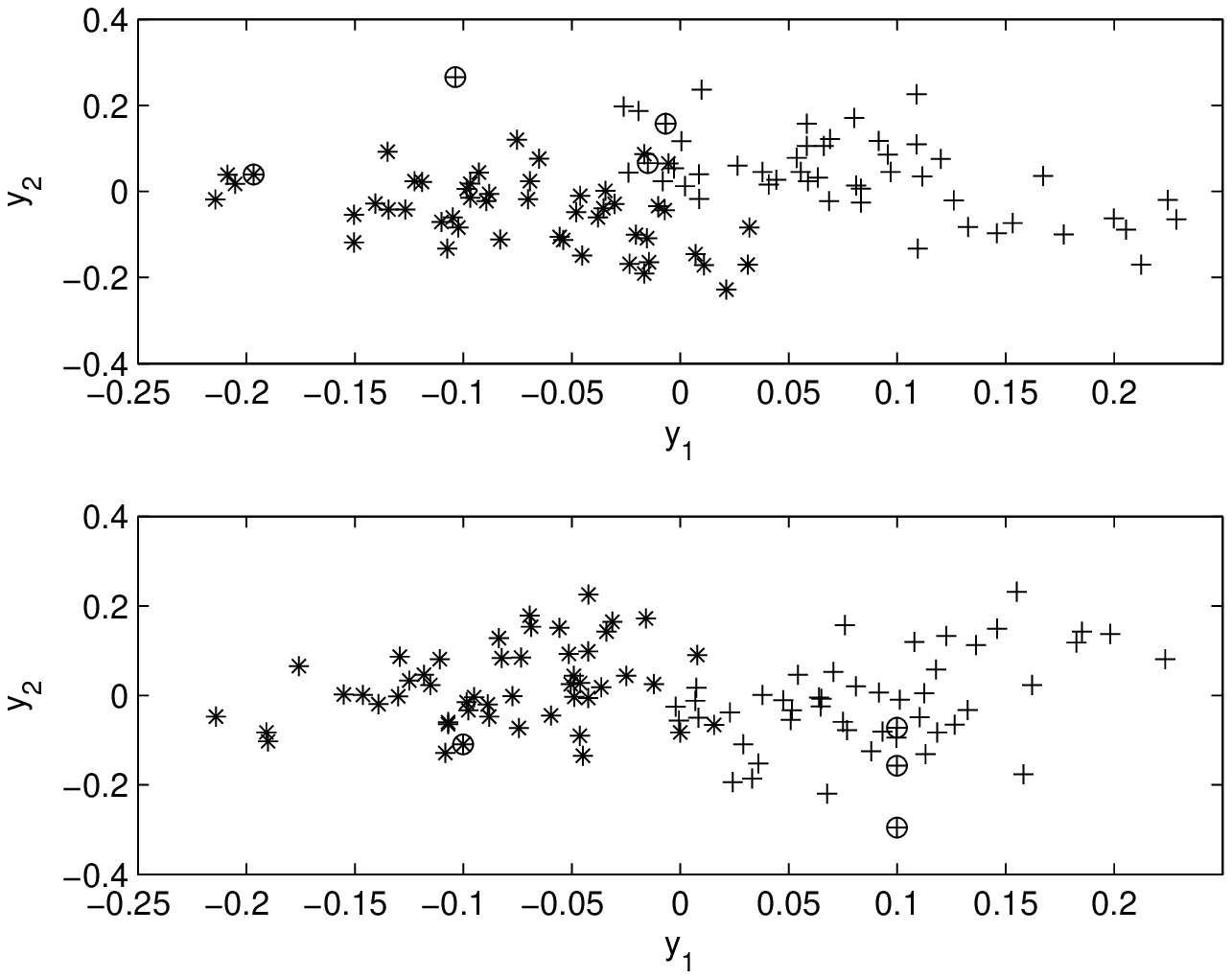,height=8.5cm}
\end{center}
\caption{{\small  (Top) The IRIS data set is depicted in the plane of the first two
principal components, $\star$ versicolor, $+$ virginica. Four selected points are
surrounded by a circle. (Bottom) The data set is represented in the plane of the first
two eigenvectors of $\mathbf{\tilde{H}}$. The linear correlation between $\bf{y^\star}$
and  the true labels is, in this case, $R=0.846$.\label{fig4}}}
\end{figure}
\begin{figure}[ht!]
\begin{center}
\epsfig{file=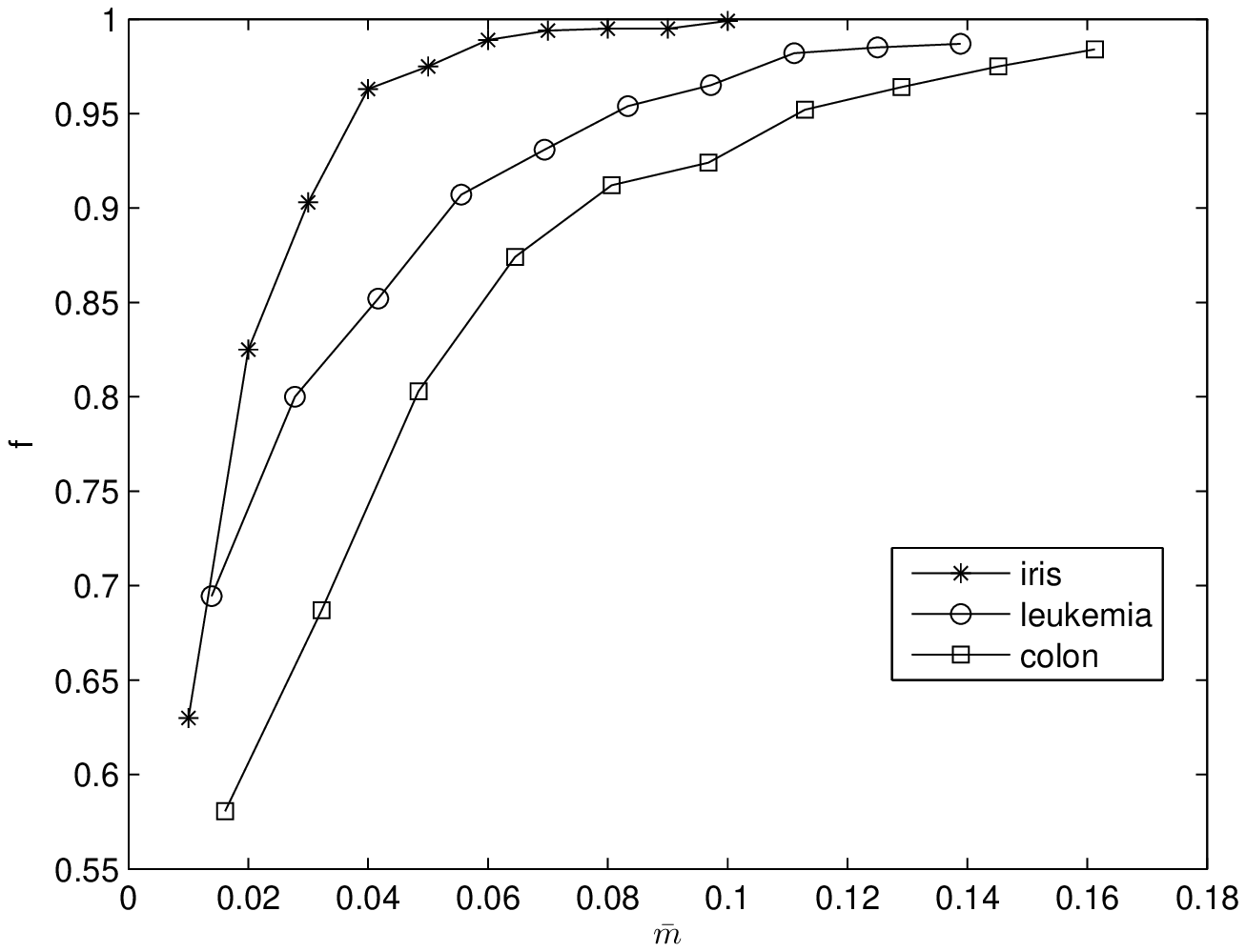,height=8.5cm}
\end{center}
\caption{{\small The fraction $f$ (see the text) is depicted as a function of $\bar{m}$
for three data sets here considered. 10000 random selections of the labeled points are
considered for each value of $m$ and for each data-set. \label{fig5}}}
\end{figure}
\end{document}